\documentclass[12pt]{article}
\newcommand{\ba}{\begin{array}}
\newcommand{\ea}{\end{array}}

\newcommand{\beq}{\begin{equation}}
\newcommand{\eeq}{\end{equation}}
\newcommand{\bea}{\begin{eqnarray}}
\newcommand{\eea}{\end{eqnarray}}
\newcommand{\beal}{\setcounter{letter}{1} \begin{eqnarray}}
\newcommand{\eeal}{\addtocounter{equation}{1} \end{eqnarray}}
\newcommand{\none}{\nonumber \\}

\newcommand{\larrow}{\,\,\,\,\hbox to 30pt{\rightarrowfill}
\,\,\,\,}
\newcommand{\slarrow}{\,\,\,\hbox to 20pt{\rightarrowfill}
\,\,\,}

\newcommand{\plabel}[1]{\label{#1}}

\def\be{\begin{equation}}
\def\ee{\end{equation}}

\def\re{(\ref }%\def\pr{(\pref}
\def\rz#1 {(\ref{#1}) }   \def\ry#1 {(\ref{#1})}
\def\el#1 {\plabel{#1}\end{equation}}
\def\rp#1 {(\ref{#1}) }

   \let\d=\delta
 \let\ep=\epsilon   
  \let\l=\lambda \let\m=\mu
\let\n=\nu    \let\s=\sigma

\let\O=\Omega \let\S=\Sigma  
\let\L=\Lambda

\def\0{\over } \def\1{\vec } \def\2{{1\over2}} \def\4{{1\over4}}
\def\5{\bar } \def\6{\partial }
\def\7#1{{#1}\llap{/}}
\def\8#1{{\textstyle{#1}}} \def\9#1{{\bf {#1}}}

\def\({\left(} \def\){\right)} \def\<{\langle } \def\>{\rangle }
 
\let\lra=\leftrightarrow 
 \let\ra=\rightarrow

\def\CL{{\cal L}}\def\CG{{\cal G}}
  
   \def\CM{{\cal M}}
 \newcommand{\dR}{\mbox{{\sl I \hspace{-0.8em} R}}}
  \newcommand{\dN}{\mbox{{\sl I \hspace{-0.8em} N}}}
\def\CP{{\cal P}}

\begin{document}

\begin{titlepage}
\renewcommand{\thefootnote}{\fnsymbol{footnote}}
\renewcommand{\baselinestretch}{1.3}
\hfill  PITHA - 96/38 
% $\quad$  ESI ... 
\\
%\medskip
%\hfill  gr-qc/96.....\\
%\medskip

\begin{center}
{\large {\bf Edge States and Entropy of 2d Black Holes}}
 \\ \medskip  {}
\medskip
%\vfill

\renewcommand{\baselinestretch}{1}
{\bf
J. Gegenberg $\dagger$
G. Kunstatter $\sharp$
T. Strobl  $\flat$
\\}
\vspace*{0.50cm}
{\sl
$\dagger$ Dept. of Mathematics and Statistics,
University of New Brunswick\\
Fredericton, New Brunswick, Canada  E3B 5A3\\
{[e-mail: lenin@math.unb.ca]}\\ [5pt]
}
{\sl
$\sharp$ Dept. of Physics and Winnipeg Institute of
Theoretical Physics, University of Winnipeg\\
Winnipeg, Manitoba, Canada R3B 2E9\\
{[e-mail: gabor@theory.uwinnipeg.ca]}\\[5pt]
 }
{\sl
$\flat$  Institut f\"ur Theoretische Physik, RWTH-Aachen\\
Sommerfeldstr. 26-28, D52056 Aachen, Germany\\
{[e-mail:  tstrobl@physik.rwth-aachen.de]}}

\end{center}

\renewcommand{\baselinestretch}{1}

\begin{center}
{\bf Abstract}
\end{center}
{\small
%\begin{abstract}
 \noindent In several recent publications Carlip, as well as Balachandran,
  Chandar and Momen, have proposed a statistical mechanical
  interpretation for black hole entropy in terms of ``would be gauge''
  degrees of freedom that become dynamical on the boundary to
  spacetime.  After critically discussing several routes for deriving
  a boundary action, we examine their hypothesis in the context of
  generic 2-D dilaton gravity.  We first calculate the corresponding
  statistical mechanical entropy of black holes in 1+1 deSitter
  gravity, which has a gauge theory formulation as a $BF$-theory.
  Then we generalize the method to dilaton gravity theories that do
  not have a (standard) gauge theory formulation. This is facilitated
  greatly by the Poisson $\s$-model formulation of these theories. It
  turns out that the phase space of the boundary particles coincides
  precisely with a symplectic leaf of the Poisson manifold that enters
  as target space of the $\s$-model. Despite this qualitatively
  appealing picture, the quantitative results are discouraging: In
  most of the cases the symplectic leaves are non-compact and the
  number of microstates yields a meaningless infinity. In those cases
  where the particle phase space is compact -- such as, e.g., in the
  Euclidean deSitter theory -- the edge state degeneracy is finite,
  but generically it is far too small to account for the semiclassical
  Bekenstein-Hawking entropy.
}%\end{abstract}

%\vfill
%\hfill November 1996  \\
\end{titlepage}

\section{Introduction}
The idea that black holes have entropy \cite{bek} has presented
theoretical physics with one of its most important and puzzling
challenges in recent years.  If it is true that we are to assign
entropy to a black hole, then according to standard wisdom of
thermodynamics, there should also be a microscopic, statistical
mechanical explanation for this entropy.  There has been considerable
progress in this direction along two very different fronts. Strominger
and Vafa \cite{strominger} have been able to derive the
Bekenstein-Hawking entropy by counting quantum mechanical string
states for the case of static, five-dimensional extreme black holes.
Their results have also been extended to a large variety of extreme
and near extremal black holes (see \cite{horowitz} for a recent
review). In all these cases, the degeneracy originates in the newly
discovered non-perturbative symmetries in the BPS spectra of string
and brane theory.

An alternative and in principle more general attempt to explain black
hole entropy has been developed by Carlip \cite{Carlip,carlip2} as
well as by Balachandran, Chandar, and Momen \cite{Balachandran}. The
main idea is roughly as follows: Classically, one is unable to find
out what is happening behind the horizon of a black hole.
Correspondingly, when quantizing, one should restrict consideration to
the part of spacetime exterior to the horizon (or 'stretched horizon'
\cite{Susskind}). This leads to the investigation of gravitational
systems on a manifold with boundary.\footnote{There are also further
  possible motivations for considering a manifold with boundary, cf.\
  \cite{Carlip, Balachandran}. One quite different alternative,
  advocated by T.\ Jacobson \cite{jacobson} as well as the school of 
  York \cite{York}, is to think of the
  boundary as an {\em outer}\/ boundary leading to a 'gravitational
  system in a box'. Another related approach is that of Maggiore, in
  which the microstates are associated with strings propogating in
  regions of spacetime near the event horizon \cite{maggiore}.} Given
a Lagrangian system with gauge symmetries on a manifold with boundary,
Carlip as well as Balachandran et al.\ argue that an appropriate
incorporation of boundary conditions leads to physical observables
that would be absent in the case of a manifold without boundary. These
'additional' degrees of freedom may be thought of as living on the
boundary, a mechanism that is also utilized in a theoretical
description of the quantum Hall effect \cite{hall, Balachandran}.
After quantization, these modes should give rise to edge states which
can in principle be traced out in order to account for the black hole
entropy.

The existence of boundary degrees of freedom can be seen most easily
in a Hamiltonian formulation \cite{Balachandran}: Gauge symmetries
give rise to (first class) constraints $G^i(x) \approx 0$ in the phase
space (here the $x$ are coordinates on the spatial part $\S$ of our
manifold $\CM = \S \times \dR$). More correctly the $G^i$ should be
smeared by some test functions (the Lagrange multiplier fields): \be
{\CG}_\epsilon := \int \epsilon_i(x) G^i(x) dx \approx 0 \, \, . \el
CG Generically the $G^i$ will include some spatial derivatives acting
on canonical variables. Suppose, for definiteness, that the $G^i$'s
are of the form \be G^i = \partial X^i(x) + F^i(X(x),...) \, \, , \el
G where $X^i(x)$ denote some canonical variables and $F^i$ are {\em
  ordinary}\/ functions of $X^i(x)$ and possibly some further
canonical variables.  To ensure functional differentiability of the
constraints ${\CG}_\epsilon$, i.e.\ to ensure well-defined Hamiltonian
vector-fields corresponding to \re{CG}), the functions $\epsilon_i(x)$
are required to vanish at the spatial boundary $\partial \S$. Thus
constraints are those functions \re{CG}) on the phase space for which
$(\epsilon_i)|_{\partial \S} =0$. Under these assumptions, phase space
function(al)s \be O_\xi := - \int \left[ X^i(x) \partial \xi_i(x) -
F^i(X(x),...)  \xi_i(x) \right] dx \el O with $(\xi_i)|_{\partial
  \S}\neq 0$ do not necessarily vanish on-shell. They coincide with
the constraints only up to boundary contributions on $\partial \S$.
Nonetheless, they give rise to a well-defined Hamiltonian vector
field. Moreover, their Poisson brackets with the constraints \re{CG})
will reflect the constraint algebra; the only difference will be the
kind of test functions appearing at the r.h.s.\ of the brackets. It
then often will happen --- and in the case of four-dimensional gravity
and the models considered below it in fact does happen --- that the
vanishing of $\epsilon_i$ on the boundary is sufficient to guarantee
that $\{O_\xi, {\CG}_\epsilon\} \approx 0$. If, furthermore, the
boundary conditions imposed on the fields do not fix $O_\xi$
completely, Eqs.\ \re{O}) give rise to non-trivial gauge invariant
(i.e.\ physical) observables, associated to the spatial boundary $\6
\S$. (Cf.\ \cite{Balachandran} for more details).

It may, however, be difficult to locate explicitly all the boundary
degrees of freedom obtained in the above manner, particularly when the
``gauge group" is actually the group of spacetime diffeomorphisms. It is for
this
reason that Carlip examined the simplified model of 2+1 Einstein gravity with
cosmological constant. This theory admits  black hole solutions discovered by
Banados, Teitelboim and Zanelli (BTZ) \cite{BTHZ}. Moreover, it has a gauge
theoretical
formulation as a Chern-Simons theory with gauge group SO(2,2) \cite{Achucarro}.
Carlip showed that if the
spacetime $\CM$ has a boundary $\6\CM$, a surface term must be added to the
usual Chern-Simons action
functional  in order to make the variational
principle well defined. For boundary conditions required so as to make $\6
\CM$ an
event horizon, the  surface term  is of the form of a
Wess-Zumino-Novikov-Witten (WZNW) $\sigma$-model coupled to external
currents  and the entropy obtained by
quantizing the boundary modes turns out to indeed coincide (to
leading order) with the semiclassical 2+1 Bekenstein-Hawking entropy
\cite{Carlip,carlip2}. Unfortunately, the quantization is not
well-understood for non-compact gauge groups like $SO(2,2)$, except in
the so-called infinite coupling limit. In his original calculation, Carlip
was able to use the fact that in this limit the WZNW theory
reduces to a bosonic string theory, in which the degeneracy of the
microstates is calculable.  In a more recent calculation
Carlip has applied this program to the Euclidean black hole in 2+1 gravity
\cite{carlip2}. In this case the gauge group is SL(2,C). Although
this is still non-compact, Carlip makes use of a result due to
Witten \cite{witten} which relates the partition function for the
SL(2,C) theory to the product of partition functions for a
corresponding SU(2) Chern-Simons theory. The
quantum mechanics  in the latter case is much better understood and
Carlip was able to derive a  result that agrees with
the one obtained in the Lorentzian case.

Attempts to generalize this calculation to the more realistic setting
of 3+1 gravity have failed thus far. The main reason for this is that the
2+1 analysis relies
very heavily on the gauge theoretic formulation of the theory, which does
not exist for 3+1 gravity.
In the present paper we intend to study the above statistical
mechanical approach to the entropy of black holes within an even lower
dimensional context, namely  1+1 dimensional gravity.  There are
at least two good reasons for doing so: First of all, it will bring
about considerable technical simplification while still allowing for a
conceptually  similar setting.  Since  the boundary of a
two-dimensional spacetime is one-dimensional, the edge degrees of
freedom will be purely mechanical ones.  Consequently the
(coupled) WZNW model in Carlip's calculation is replaced by a point
particle model with a finite dimensional phase space. The
mathematics involved is therefore more straightforward, allowing one
to focus on  conceptual rather than technical issues.

The second motivation for considering 2-D gravity concerns the
universal validity of Carlip's approach.  As Wald's work emphasizes
\cite{wald}, the classical thermodynamical behaviour associated with
event horizons can be derived in a model-independent way. Only the
details differ from one theory to another. This raises the following
important question: Do edge states account for entropy in {\em any}\/
geometrical theory that admits solutions with event horizons?
Although we will not be able to answer this question definitively, we
will be able to apply Carlip's method to a large class of $\mbox{2-D}$
gravity theories. In this context it is particularly important to note
that almost all of the $\mbox{2-D}$ models considered below do {\em
  not}\/ allow for a formulation in terms of an ordinary gauge theory.
We are nonetheless able to generalize and apply the methods developed
for theories such as 2+1 gravity that do have a gauge theoretic
formulation. These models therefore provide the first examples of
non-gauge theoretic models where the approach of Carlip and
Balachandran et al. may be applied and tested.

Somewhat surprisingly, we do {\em not}\/ get the expected results for
any of the models considered.  It is not clear at this stage whether
this is due to the special features (i.e.\ finite number of degrees of
freedom) of the edge dynamics in 2-D, or whether it indicates that the
general method lacks universal validity in its present form.

The theories we wish to consider include the general class of dilaton gravity
models governed by a Lagrangian for a 2d metric $g$ and a dilaton
field $\Phi$ of the form \cite{generic}
\begin{equation}
L_{\mbox{gdil}}[g,\Phi] = \int_M d^2 x \sqrt{|\det g|} \left[D(\Phi) R -
V(\Phi) + Z(\Phi) g^{\m\n} \6_\m \Phi \6_\n \Phi \right] \, .
\plabel{L}
\end{equation}
Our considerations will be applicable to {\em any}\/ Lagrangian of the
above form (provided that the functions $D,V,Z$ are smooth, $D' \neq
0$ and either $Z \neq 0$ or $Z \equiv 0$).  However, for simplicity we
will mostly restrict ourselves to the case $D=\Phi$ and $Z \equiv
0$.\footnote{Actually this is not much of a restriction: Under the
  above assumptions this form of $L_{\mbox{gdil}}$ may be obtained
  always by a redefinition of the basic fields (e.g., $\Phi \ra
  D(\Phi)$) \cite{domingo1}.} The deSitter model results from \re{L})
by further choosing $V \propto \Phi$.  Only in this particular
case, as well as for $V=const$ (reformulated string inspired gravity),
the theory \re{L}) (with $D=\Phi$ and $Z \equiv 0$) may be
reformulated entirely in terms of a standard gauge theory
\cite{Isler,cangemi}. For other choices of the 'potential' $V$ we have
to leave the realm of ordinary gauge theories, and, consequently, from
this perspective we are closer in spirit to the 4d theory than in
Carlip's 2+1 treatment.  It should also be noted that \re{L})
incorporates the spherical reduction of the Einstein Hilbert action
\cite{spher}. Thus our treatment should also cover the {\em
  four}\/-dimensional Schwarzschild black hole.\footnote{With a slight
  generalization of \re{L}) we may also treat charged black holes.
  Only for rotating black holes other techniques would be required.}
Let us remark, however, that in the present context the suppressed
dimensions could become vital; the handling of the boundary may break
the rotational invariance (or additional fields may be needed to
restore it). Finally, we note that using the general considerations of
Wald \cite{wald}, or a straightforward analysis specific to the
present class of models \cite{domingo2}, it can be shown that for
solutions to \re{L}) with event horizons, the thermodynamical entropy
is generically given by \be S_{thermo} = {4\pi D(\Phi_0)\0 \hbar}
\label{thermo_entropy}
\ee where $\Phi_0$ is the value of the dilaton field at the outer
horizon.  At the end of Section 2 one of several  derivations of
\re{thermo_entropy}) will be recapitulated briefly.

The specific model closest in spirit to Carlip's original calculation is
that of deSitter gravity, for which $V\propto \Phi$. This theory can
be written in gauge theoretic form as a BF topological field theory
\cite{Isler} with Lie algebra so(2,1). Moreover, it can be obtained by
dimensional reduction (imposing axial symmetry) on 2+1 gravity. Thus,
the black hole solutions in this model are the dimensionally reduced
versions of the BTZ black hole \cite{Ortiz}.  We will show that the
boundary action of the deSitter model turns out to be that of a
coadjoint orbit. The quantization of such a system is standard and
straightforward. We provide a brief review of various methods for
doing this in an Appendix, including a method recently developed
\cite{Proceedings} for the case of Euclidean black holes in 2-D
deSitter gravity, in which the system can be quantized in terms of a
system of two harmonic oscillators with fixed total energy.
Quantization in this case yields a discrete mass spectrum and the
degeneracy of states turns out to be proportional to the square root
of the mass. The resulting entropy of $\ln{\sqrt{M}}$ disagrees with
the Bekenstein-Hawking value, which is proportional to
$\sqrt{M}$.\footnote{We note, however, that the logarithmic
  result is in  qualitative  agreement with a numerical  investigation
  of the entanglement entropy of 1+1 dimensional gravity systems
  performed by Srednicki \cite{Sred}.}

As previously noted in the case of general 2-D dilaton
gravity a gauge theoretic formulation does not exist.  The main tool
for treating the general class of models \re{L}) will thus be their
equivalent formulation as so-called Poisson $\s$-models
\cite{ModLetts} (cf.\ also \cite{Brief,PartI}). From a certain
perspective the latter are {\em non-linear}\/ gauge theories
\cite{Nieuwenhuizen,Ikeda}. For us it will be more important, however,
to regard them as $\s$-models of a very special kind: While spacetime
is the worldsheet of the model, its target space is a 'Poisson
manifold'.  This means that the target space of the theory, which is
just $\dR^3$ in the present context, carries a Poisson bracket.  This
Poisson bracket is degenerate. On restriction to appropriate
two-dimensional submanifolds (of the $\dR^3$), the bracket becomes
non-degenerate; thus these submanifolds are symplectic and the full
three-dimensional target space foliates into (generically)
two-dimensional symplectic leaves. It turns out that a target space
coordinate (i.e.\ a certain function of the fields of the $\s$-model),
which may be used to label the different symplectic leaves, may be
identified on-shell with the mass $M$ of the black holes of the
classical solutions.

The knowledge of this somewhat unexpected structure has been used in
several works to efficiently solve the general models \re{L}) at both
the classical and quantum level \cite{ModLetts}. In the present context,
this hidden Poisson structure will become even more important. On our
(spacetime) manifold with boundary we will fix boundary conditions so
as to pick out a black hole of fixed mass $M$. We will then derive the
point particle action induced at the boundary following the strategy
of \cite{Carlip,Balachandran}. The phase space of this point particle
will turn out to be {\em identical}\/ to the symplectic leaf singled
out by the choice of $M$!  So, whereas in previous works the Poisson
structure was used for auxiliary purposes in solving \re{L}), now the
symplectic leaves of the target space become 'alive': they provide the
phase space for the edge dynamics.  Thus, the final picture that
emerges for general 2-D dilaton gravity is very appealing.
Still, it does not seem to yield reasonable results for the
statistical mechanical entropy.

The paper is organized as follows: In Section 2 we review the logic of
\cite{Carlip,Balachandran} for obtaining the WZNW action governing the
boundary degrees of freedom in the 2+1 theory.  In the process, we
point out some puzzling conceptual features of the approach, which are
then made even more explicit when applying the procedure to 2d
deSitter gravity.  In Section 3 the discussion is extended to the case
of general 2-D dilaton gravity. Section 4 concludes with a short
summary and  outlook.

\section{Gauge-Theory-like Gravity Models}

As shown in  \cite{Achucarro}, 2+1 gravity, with or
without a cosmological constant $\L$, may be formulated in terms of
a Chern-Simons gauge theory \be CS[A] = \frac{k}{4 \pi} \int_{\CM} tr
\left( A \wedge dA + \frac{2}{3} A \wedge A \wedge A \right)\,\,.
\el CS Here $A$ is a standard gauge field, i.e.\ a Lie-algebra
valued one-form. For $\L \neq 0$ the appropriate gauge group is
$SO(2,1) \times SO(2,1)$ for Lorentzian signature and $SL(2,C)$
for the Euclidean signature of the gravity theory.  So $A$ is
a collection of six one-forms, three of which may be
identified with the threebein of the gravity theory, while the
remaining ones coincide with the three components of the
spin-connection.

As required for a gravity theory, \re{CS}) is invariant under
diffeomorphisms. If $\CM$ has no boundary, \re{CS}) is also invariant
with respect to non-abelian gauge transformations (connected to the
identity). To see this explicitly, we perform a gauge transformation
$A^g= g^{-1} A g + g^{-1} d g$: \be CS(A^g) = CS(A) + \frac{k}{4 \pi}
\int_{\6 \CM} tr ( A \wedge dg g^{-1} ) - \frac{k}{12 \pi} \int_{\CM}
tr (dg g^{-1} \wedge dg g^{-1} \wedge dg g^{-1} ) .  \el CSg The
second term on the r.h.s.\ vanishes if $\CM$ has no boundary or if $g$
vanishes at $\6 \CM$. The third term also yields a boundary
contribution only, as long as $g(x)$ is connected to the identity map
in the gauge group. For arbitrary 'gauge transformations' $g(x)$ the
CS-action picks up multiples of $2 \pi k$ in addition to boundary
terms; however, for the standard choice $k$ an integer (times $\hbar$)
such contributions do not contribute to the path integral.

It is important to note that the diffeomorphism symmetry of \re{CS})
is not independent from its non-abelian gauge symmetry.  In fact one
may easily verify the following identity \cite{Achucarro} \be \CL_v A
= d(v^\m A_\m) + [ A, v^\m A_\m] + \mbox{field equations} \, \, , \el
diff where $\CL_v$ denotes the Lie derivative of the vector field $v$.
This equations shows that {\em on-shell}\/ an infinitesimal
diffeomorphism may be generated by a gauge transformation with $g=1 +
v^\m A_\m + ...$, where the dots indicate terms of higher order.
Correspondingly, at the Hamiltonian level of the theory there will be
no independent additional first class constraints generating
diffeomorphisms. Thus, in a theory like \re{CS}) it is sufficient to
consider non-abelian gauge symmetries; diffeomorphisms (as well as
frame-bundle rotations of the vielbein) are taken care of
automatically.

We now wish to consider the case $\6 \CM \neq 0$.  It is therefore
necessary to fix some boundary conditions on $\6 \CM$.  Introducing an
auxiliary complex structure on this two-dimensional surface, let us,
for concreteness, fix the $z$-component of $A$: $(A_z)|_{\6 \CM}: =
\alpha$. With the addition of a boundary term $B[A] \equiv
(k/4\pi)\int_{\6 \CM} A_{\bar z} A_z d \bar z \wedge d z${}$\,${}the
variation of $CS[A]$ will have no unwanted boundary contributions.
However, the total action $CS[A] + B[A]$ is not invariant under gauge
transformations that do not vanish at $\6 \CM$.  This can quite easily
be corrected by introducing group-valued fields $g$ living at the
boundary.\footnote{Here we use an old trick: If an action behaves like
  $I(A^g) = I (A) + J(A,g)$ under gauge transformations $A \ra A^g$,
  then clearly $I(A) + J(A,g)$ will be invariant under the
  simultaneous gauge transformations $A \ra A^h$ and $g \ra h^{-1}
  g$.} The total gauge-invariant action then takes the form
\begin{equation} \widetilde{CS}[A,g] = CS[A] + WZNW[g] + \frac{k}{4
    \pi} \int_{\6 \CM} tr \left(2 A_z \6_{\bar z} g g^{-1} + A_{\bar
    z} A_z \right) d^2z \,\,, \label{CSnew} \end{equation} where
  $WZNW[g]$ is the action of the standard Wess-Zumino-Novikov-Witten
  model \cite{WZNW}\footnote{The somewhat unconventional minus sign
    between the two terms may be changed by passing from $g$ to
    $g^{-1}$.}: \be WZNW[g] = \frac{k}{4 \pi} \int_{\6 \CM} tr \left(
  \6_z g g^{-1} \6_{\bar z} g g^{-1}\right) d^2z - \frac{k}{12 \pi}
  \int_\CM tr \left( dg g^{-1} \right)^3 \, \, .  \el wznw According
  to Carlip and Balachandran {\it et al}, the $g$-dependent part of
  the action, i.e.\ the boundary action, must be quantized. The
  degeneracy of the resulting quantum theory should account for the
  statistical mechanical entropy of the black hole.
\par
There are in fact several ways of deriving the above boundary action.
The one chosen here, which followed primarily \cite{Balachandran}, is
probably the shortest one. It is not, however, particularly
instructive from a conceptual viewpoint. The reason is that we added
the $g$-dependent terms in order to make the action gauge invariant.
However, the WZNW-part of the action would {\em not}\/ be needed to
restore gauge invariance under {\em those}\/ transformations that
respect the boundary condition $(A_z)|_{\6 \CM} = \alpha$. One can
therefore avoid the need for introducing edge dynamics by simply
restricting to gauge transformations that respect the boundary
conditions.
\par
We will therefore sketch one further derivation of the boundary action
in \re{CSnew}).
Again we start with the CS-action \re{CS}).  Now, however, we fix
boundary conditions for $A_z$ only {\em up to gauge
  transformations}.\/ To be more specific, we parametrize $A$ in terms
of a quantity $\bar A$ that is fixed at the boundary and an
unrestricted group-valued variable $g$: \be A= g^{-1} \bar A g +
g^{-1} d g \, \, . \el barA Inserting this new parametrization into
\re{CS}), we end up with the r.h.s.\ of Eq.\ \re{CSg}) with bars on
top of all the $A$'s.  For a well-defined variational problem we need
to add a boundary term to kill the $\delta \bar A_{\bar z}$
contributions at $\6 \CM$. For this purpose we could add, e.g.,
$(k/4\pi)\int_{\6 \CM} tr \left( \bar A_{\bar z} (\bar A_z + \6_z g
g^{-1}) \right)d^2z$.  If, however, we require that everything added
be expressible in terms of the original variables $A$ alone, i.e.\ in
terms of the combination \re{barA}) of $\bar A$ and $g$, then the
added term becomes unique and we end up with \re{CSnew}) with all the
$A$'s replaced by the gauge-fixed $\bar A$'s. Note that by
construction of \re{CSnew}) this is equivalent to
$CS[A]+(k/4\pi)\int_{\6 \CM} tr (A_z A_{\bar z} )d^2z $ (without
bars!) after using \re{barA}).

It is important to keep in mind that this second derivation leads to
a significant interpretational shift, since we are now fixing $\bar
  A$ at the boundary instead of $A$ (which is equivalent to saying
  that we fix $A$ up to a gauge transformation).  Choosing boundary
  conditions such that the 'threebein-components' of $\bar A$
  correspond to a (possibly stretched) horizon, the
  threebein-components of $A$ will for general $g$ not correspond to a
  horizon.\footnote{It may be difficult to locate this in {\em
      explicit}\/ coordinate calculations in the present context. This
    comes about since the BTHZ 'black hole' is just a quotient space
    of 2+1 (anti-)deSitter space by some (properly discontinuously
    acting) discrete subgroup. As such any {\em local}\/ patch looks
    completely alike. A 'horizon' can be defined only taking into
    account some nontrivial global issues.  This changes drastically
    in the 1+1 models considered below (or in the following Section).
    There the (Killing) horizon may be read off from the {\em local}\/
    value of the dilaton field and the above statement becomes
    completely evident in that case.} Since $A$ is to be identified
  directly with the metrical variables, by adopting this second
  derivation we have to give up the picture that $\6 \CM$ corresponds
  on-shell to a (stretched) horizon (of the classical spacetime
  solution). From this perspective, the variables $g$ now parametrize
  the 'deviation' of $\6 \CM$ from the horizon, and this can become
  arbitrarily large now. Perhaps this is not unnatural: the relevant
  modes no longer ``live on the boundary'', but instead correspond to
  vibrational modes of the boundary itself.

This last derivation is in spirit very close to the one of
Carlip.\footnote{And we gratefully acknowledge discussions on it with
  him.} However it is important to note that in order
for this derivation
to work, one must give up the notion that $\partial M$ corresponds
to the horizon on-shell. This point is not made clear in Carlip's original
paper \cite{Carlip}.
In fact, if one tries to implement Carlip's program without letting
the location of the boundary ``fluctuate off the horizon'',
 one runs into the following problem: First fixing
boundary conditions for $A$, (e.g. fixing $A_z$ at $\6\CM$), then
decomposing $A$ according to \re{barA}) and finally treating
$\bar A$ as background currents that are 'fixed by the boundary data',
would force the variable $g$ to  be restricted at $\6\CM$ too. For
example, if $\bar
A_z$ is constant on $\6\CM$, then  according to \re{barA}) $g$ may depend on
$\bar z$ but not on $z$ and the WZNW-action would vanish trivially.
In other words: Boundary data do indeed restrict the gauge symmetry at
the boundary. However, by their nature they certainly also eliminate
some of the original fields at the boundary (namely those fixed by the
boundary data). In the present case these two mechanisms balance (the
boundary data are a good cross section to the gauge symmetries at the
boundary).

Irrespective of how one motivates or interprets the boundary
contributions in \re{CSnew}), there still is another, mathematical
question: The constraints following from \re{CS}) are of the form
\re{G}). It is not clear at this stage that the observables \re{O})
are indeed in one-to-one correspondence with the group elements $g$
introduced as the edge degrees of freedom in \re{CSnew}), as %is
assumed always in the literature.  %In fact, below we will encounter a
%similar two-dimensional situation, where an analogous correspondence
%does in fact {\em not}\/ hold.

Taking the $g$-dependent part of \re{CSnew}) as his starting point,
Carlip \cite{Carlip} derives an entropy for the 2+1 black hole
\cite{BTHZ} by counting quantum states in the edge model. In his
earlier works this has been done for the Lorentzian signature of the
metric. What he actually counted was rather the number of irreducible
representations of the appropriate current algebra. In his latest work
\cite{carlip2} Carlip redid the calculation for Euclidean signature,
for which the theory can be formulated in terms of a compact gauge
group and the quantization of the coupled WZNW model is therefore
better defined. A key feature of Carlip's calculation was the need to
impose a constraint on physical edge states. This constraint was
roughly a remnant of the Wheeler-DeWitt equation for the bulk theory,
and in the case of Lorentzian calculation at least, was needed to
reduce the number of edge states to a finite number.
In both the Lorentzian and Euclidean
regimes, Carlip found the resulting entropy to coincide to leading
order in $\hbar$ with the semiclassical Bekenstein-Hawking formula
obtained by thermodynamical approaches \cite{TeitelCarlip}.

Let us now turn to the simplest two-dimensional model, deSitter
gravity \cite{deSitter}.  This model, characterized by $D=\Phi$, $V
\propto \Phi$, and $Z \equiv 0$ in \re{L}), may be formulated in terms
of a non-abelian gauge theory of the BF-type \cite{Isler}: \be
L_{\mbox{deS}}[A,B] = -2 \int_\CM tr (BF) \, \, . \el BF Here $F=dA+A
\wedge A$ is the curvature two-form of a standard gauge field $A$,
while $B$ is a Lie-algebra valued function on two-dimensional
spacetime $\CM$.  The appropriate Lie algebras are $so(2,1)$ and
$su(2)$ for Lorentzian and Euclidean signature of the metric theory,
respectively (in the latter case this only holds only for an appropriate
sign of the proportionality constant in $V$). Thus, e.g., $B=B^i T_i$
where $T_i$ are generators of the respective Lie algebra, normalized
such that $-2 tr( T_i T_j ) = \delta_{ij}$ in the case of $su(2)$,
while for $so(2,1)$ the normalization of $T_1$ is altered to $+2tr(T_1
T_1)=1$.  The identification of the metric variables with the gauge
theory variables is similar to the one encountered in the
2+1-dimensional theory: The first two components of $A$ are identified
with the zweibein while the last component corresponds to the spin
connection.  The first two components of $B$, on the other hand, are
Lagrange multipliers enforcing torsion zero and $B^3$ turns out to be
the dilaton $\Phi$. The action of diffeomorphisms on $A$ is again as
in Eq.\ \re{diff}).  An analogous equation may also be proven for $B$
so that the non-abelian gauge transformations again generate
diffeomorphisms on-shell.

Next we determine the boundary action, following the simplest route.
First let us agree to fix boundary conditions for $B$ and not for $A$
(for some motivation of this see below). Fixing
$B$ (entirely or up to gauge transformations, as one prefers, cf.\ the
discussion above), we should add a boundary term of the form
$-\int_{\6 \CM} tr(BA)$. To make this gauge invariant with respect to
gauge transformations restricted in no way at the boundary, we again
should add group-valued edge variables. The appropriate action then is
of the form
\be \widetilde{L_{\mbox{deS}}}[A,B,g] = -2 \int_\CM tr (BF) + 2
\int_{\6 \CM} tr \left(B (A + dg g^{-1})\right) \, \, , \el deSnew
which clearly is invariant under
\be A \ra A^h = h^{-1} A h + h^{-1}   d h \, \, , \;\; B \ra B^h =
h^{-1} B h \, \, , \;\; g \ra g^{h} =   h^{-1}g \, \, \el Bg
for arbitrary maps $h(x)$ from $\CM$ to $G$.

Before turning to the quantization of the boundary action, some
remarks are in order: First, {\em on-shell}\/ the two components $B^a$
of $B$ coincide with a Killing vector field $k^a$ in an orthonormal
frame bundle basis.  Actually a similar statement holds for all the
theories \re{L}), which locally always have (at least) one Killing
field $k$ (cf., e.g., \cite{domingo2,PartI}). The horizon of a black
hole thus coincides with lines where $(k)^2 \equiv B^a B_a =0$.  For
Lorentzian signature of \re{L}), the horizon may therefore be
characterized by $B^+ \equiv (B^1 + B^2)/\sqrt{2} = 0$ and/or $B^-
\equiv (-B^1 + B^2)/\sqrt{2} = 0$. The bifurcation point where the
past and future horizon of an eternal black hole coincide is specified
by $B^+=B^-=0$, which is a hyperbolic fixedpoint of $k$ in this case.
For Euclidean signature, on the other hand, the horizon degenerates to
a point, $B^1=B^2=0$, which now is an elliptic fixed point of $k$.

The black hole mass, furthermore, may be identified, on-shell, with the
gauge-invariant combination \be M= -2tr(B \cdot B) \equiv B^a B_a +
(B^3)^2, \el casimir where $B^a B_a$ equals $2B^+B^-=-(B^1)^2+(B^2)^2$
in the Lorentzian and $(B^1)^2+(B^2)^2$ in the Euclidean theory,
respectively.\footnote{The appropriate generalization of this for the
  general nonlinear gauge theory \re{L}) will be dealt with in the
  next section.} Thus at the horizon $B^1=B^2=0$ and $B^3 \equiv \Phi
= \sqrt{M}$, while at a 'stretched horizon' the $B^i$ would be close
to these values.

This simple option to enforce ones boundary to coincide with a
(stretched) black hole horizon may be seen as a motivation for
choosing to fix $B$ at the boundary $\6 \CM$. In this context it is
then also particularly clear that if we adopt the viewpoint that $B$ at
$\6 \CM$ is fixed only {\em up to gauge transformations}\/ (i.e.
$B_{\6\CM} = (g^{-1} \bar{B} g)^3$ with $\bar{B^a}=0$ and
$\left(\bar{B^3}\right)|_{\6 \CM} = \sqrt{M}$) then $(B^3)|_{\6 \CM}$
may take any value depending on the choice of $g$ and the boundary
will be a horizon only for $g$ in the stabilizer $H$ of the Lie
algebra element $T_3$. In this picture $g \in G/H$ is found to
precisely determine the location of (a point on) $\6 \CM$ (at a given
time) within the boundaryless maximal extension of $\CM$. We believe
that this feature is not an artifact of the two-dimensional model at
hand, but that it is generic and present also, e.g., in the 2+1
theory, only there it is less obvious due to the lack of $B$-fields
(cf.\ also the remarks in footnote 8).

Note also that if we fixed boundary conditions merely for the $A$'s,
we would induce {\em no}\/ boundary terms in \re{BF}), which moreover
is already gauge invariant. On the other hand, the constraints would
be of the form \re{G}) and observables of the form \re{O}) should
exist. A similar discrepancy may be observed also in the case of an
abelian  $BF$-action even when treating boundary conditions for 
 the $B$-field (cf.\ also the remarks in the Outlook).

Following \cite{Carlip,Balachandran}, we now need to quantize \be
L_{\mbox{coad}} = 2 \int_{\6 \CM} tr \left(B dg g^{-1} \right) \, ,
\el coad where $B$ should be thought of as a fixed element in the Lie
algebra.  This is quite straightforward and was solved some time ago
\cite{KostantSoriau,Woodhouse}. To keep this work self-contained and
to keep the mathematical aspects of the 2-D model as transparant and
simple as possible, we now give a brief account of the quantization of
\re{coad}), referring to the Appendix for further details.

We start with a simultaneous treatment of the Lorentzian and Euclidean
signature, corresponding to $G=\widetilde{SO_e}(2,1)$ (cf.\ last
reference in \cite{Isler}) and $G=SU(2)$, respectively, where
$\widetilde{SO_e}(2,1)$ denotes the universal covering group of the
component of $SO(2,1)$ connected to the
identity.\footnote{$\widetilde{SO_e}(2,1)$ has no finte-dimensional
  faithful matrix representation; so the trace in \re{coad}) is to be
  replaced by the Killing metric in this case.} Furthermore let us
align $B$ with the three-direction in the Lie algebra: $B=\sqrt{M}
T_3$, $M$ constant.  Any other choice of boundary conditions for $B$,
at least if they are constant along $\6 \CM$ and if $M>0$ in
\re{casimir}) certainly, may be mapped to this problem by an
appropriate change of variables in \re{coad}). Thus the point particle
action $L_{\mbox{coad}}$ reduces to \be L_{\mbox{coad}}=2\sqrt{M}\int
tr(T_3 \dot g g^{-1}) dt \el Lcoad where $t$ denotes the coordinate
along $\6 \CM$. This action is already in first order form and
basically coincides with an action $\int p \dot q dt$. The Hamiltonian
of the system is zero (strongly, not only weakly), as it should be for
a diffeomorphism invariant system. (Note that we have not broken the
diffeomorphism invariance along the boundary; this could occur only,
if the boundary conditions for $B$ were chosen to be explicitly
time-dependent).  It still remains to determine the phase space
topology (space of values for $q$ and $p$), which will turn out to be
nontrivial.

Since $g$ takes values in a three-dimensional group $G$, the phase
space can be at most of dimension two. Indeed, parametrizing $g$
according to $g= l \, \tilde g$, where $l=\exp(\lambda T_3) \in H$,
$H$ being the one-dimensional stabilizer subgroup of $B \propto T_3$,
so that $\tilde g \in G/H$, we find \be L_{\mbox{coad}} = 2\sqrt{M}
\int \left[ tr \left( T_3 \dot{\tilde g} \tilde g^{-1} \right) - \2
\dot \lambda \right] dt \, \, .  \el coad2 So, the one dimension along
$H$ drops out from the action as a total divergence. The phase space
of the point particle at the boundary is found to be $G/H$, endowed
with the (Kirrillov) symplectic two-form
$$\Omega = 2 \sqrt{M} \, tr \,
\left( T_3 \, d\tilde g \tilde g^{-1} \wedge d\tilde g \tilde g^{-1}
\right)$$.

In order to complete the calculation, we must now specify the signature. We
consider the Euclidean theory first. In this case
$G=SU(2)$, $H=U(1)$ and $G/H = S^2$ (the famous Hopf fibration of
$SU(2)\sim S^3$). We can parametrize $G$ as
\be g= \displaystyle{ \left(
\begin{array}{cc} \cos{\frac{\theta}{2}} \; e^{i(\varphi - \l)/2} &
  \sin{\frac{\theta}{2}} \; e^{-i(\varphi + \l)/2} \\ \; & \; \\
  -\sin{\frac{\theta}{2}} \; e^{i(\varphi + \l)/2} &
  \cos{\frac{\theta}{2}} \; e^{-i(\varphi - \l)/2} \end{array}
\right)} \el g with $\theta \in [0,\pi]$, $\varphi\, \in [0,2\pi]$,
and $\l \in [0,4\pi]$. Here $\theta$ and $\varphi$ become the standard
spherical variables of the phase space $S^2$, while $\l/2$ is the
angular variable along the diagonal subgroup that was found to drop
out.\footnote{We are using the representation $T_i=-i\s_i/2$, where
  $\s_i$ are the standard Pauli matrices.} Furthermore, it is
straighforward to verify that $p:= \sqrt{M}\cos \theta$ and $q :=
\varphi$ provide Darboux coordinates, i.e.\ in these coordinates
$\Omega = dp \wedge dq$.  Certainly this equation holds locally only,
everywhere at the two-sphere except for its poles $p=\pm \sqrt{M}$;
$\Omega$, being proportional to the volume-form, is closed but not
exact.

A system with a compact two-dimensional phase space may be quantized
only if $\oint \Omega = 2 \pi n \hbar$, $n \in Z${}
$\big($(generalized) Bohr-Sommerfeld quantization condition, cf.\ also
the Appendix and, e.g., \cite{Woodhouse}$\big)$.  This yields the
consistency condition: \be M = n^2 \hbar^2/4 \, \, . \el quant Thus,
it is only for these values of the mass $M$ (up to a possible shift,
cf.\ the discussion in the Appendix) that a consistent quantization of
the boundary action can be carried out. The constraint \re{quant}) may
also be obtained by requiring \re{coad}) to become single-valued on
$G/H \sim S^2$ up to multiples of $2\pi \hbar$. In fact the action
$L_{\mbox{coad}}$ is of the Wess-Zumino type, analogous to the second
term in \re{wznw}), an observation that will become more transparent
in the formulation of the following section. Here in view of
\re{coad2}) this may be verified by noting that $\lambda/2 \sim
\lambda/2 + 2 \pi$, which again enforces \re{quant}) for consistency.

In the above the parameter $M$ entered as a classical quantity,
coinciding with the mass of the classical spacetime solution.  If, on
the other hand, we quantize the gravitational field by quantizing the
$BF$ {\em bulk}\/ action, $M$ becomes an operator on the space of
physical wavefunctionals $\Psi[A]$ or $\Psi[B]$. It is therefore
reassuring that the spectrum of the corresponding quantum operator $M$
coincides with the one found in \re{quant}) (up to possibly precisely
the shift(s) discussed in the Appendix, cf.\ the ongoing discussion on
the issue of the spectrum of $tr (B \cdot B)$ in, e.g.,
\cite{Hetrick}): In the connection representation $\Psi[A]$, gauge
invariance of the physical wave functionals restricts $\Psi$ to live
on the Weyl cell of $su(2)$, which is just an interval along $T_3$.
The Casimir $M \propto tr(B)^2$, on the other hand, becomes nothing
but the Laplacian on the Weyl cell; clearly here (and similarly for
other compact gauge groups) this yields a discrete spectrum for $M$.

The next step in Carlip's program is to determine the number $N$
of quantum states for the edge degrees of freedom living on the
boundary of a spacetime with mass $M$ satisfying \re{quant}). Clearly
this will be finite because  the phase space is compact.  This is quite
satisfactory since no regularization will be required when
calculating the statistical mechanical entropy via $S_{\mbox{st.m.}}= k
\ln N$.

An approximate answer for the degeneracy of states can be obtained
immediately. From elementary statistical mechanics we know that there
will be about one quantum state per phase space volume $h$. Since the
total symplectic volume was found to be $n h$, this leads to $N \sim
n$.  In the Appendix we present three precise calculations by applying
three different quantization procedures to the present system. Two of
these lead to $N=n$, one to $N=n+1$. This difference is not
crucial for our purposes, and we will stick to $N=n$ in what follows.
Thus we may conclude \be S_{\mbox{st.m.}}\propto \ln n = \ln (2
\sqrt{M}/\hbar) \, \, .  \el SdeS

Unfortunately this result does not match the thermodynamical entropy for this
model using other methods \cite{domingo2,wald}.
For example, Wald's general method starts with the variation of the action
under spacetime diffeomorphisms of the
form:
\bea
x^\mu\to x^{'\mu}&=&x^\mu+\delta x^\mu\none
\Phi^A(x)\to \Phi'{}^A(x') &=& \Phi^A(x) + \delta \Phi^A(x)\quad ,
\eea
where for the moment we use a condensed notation in which the complete set of
fields (including
the metric) is denoted by $\Phi^A(x)$, and $x$ is the spacetime coordinate.
Under such a general  transformation,  an  action which is second order in
derivatives of the fields
has the following variation:
\be
\delta I = \int d^2x \left( {\delta I\over \delta \Phi^A} \delta \Phi^A
+ {\partial j^\mu \over \partial x^\mu}\right) \quad ,
\label{eq: delta I}
\ee
where $j^\mu$ is the associated Noether current.
Diffeomorphism invariance of the action requires that the Noether current be
divergence free when
the classical field equations
are satisfied.
\par
For the action \re{L})  in a parametrization for which $D(\Phi) = \Phi$ and
$Z(\Phi) =0$, the Noether current is \cite{domingo3}:
\bea
j^\lambda &=&   (\Phi R - {V}) \delta x^\lambda   - \nabla_\sigma\Phi (
g^{\alpha\lambda} g^{\beta\sigma} - g^{\alpha\beta}
g^{\lambda\sigma} ) \bar{\delta} g_{\alpha\beta}\none
 & & +  \Phi( g^{\alpha\sigma} g^{\beta \lambda} - g^{\alpha\beta}
g^{\sigma\lambda}
) \nabla_\sigma(\bar{\delta} g_{\alpha\beta}) \quad ,
\label{eq: noether 1}
\eea
where $\bar\delta$ denotes variation of the corresponding field under Lie
derivation along $\delta x^\mu$.
As remarked already above, all the solutions have a Killing vector
$k$. One can show that it may be written as
\be k^\lambda = - \epsilon^{\lambda\sigma}
\nabla_\sigma \Phi \, \, . \el Killing
Choosing  $\delta x^\lambda := - k^\lambda$,
the variations of the scalar field and metric will vanish on-shell for such
transformations, and the Noether current becomes:
\be
j^\lambda = \left({\Phi}  {dV \over d\Phi} -
{V}\right) \epsilon^{\lambda\sigma} \nabla_\sigma\Phi\none
\label{eq: noether 2}\quad ,
\ee where we made use of the field equation $R=V'(\Phi)$.  Clearly the
equation $\6_\m j^\m=0$ implies that the Hodge dual of $j$, $J_\mu =
\epsilon_{\mu\lambda} j^\lambda$ is closed (on-shell) and thus locally
exact: \be J=(V-\Phi V'(\Phi)) d\Phi = dQ \quad . \el J By means of
the field equations one may verify, furthermore,  that the Noether charge $Q$
associated with the Killing vector is:
\be Q = V \Phi - g^{\alpha\beta}
\nabla_\alpha\Phi\nabla_\beta \Phi \,\,.  \ee

According to Wald's prescription, the black hole entropy should be
$2\pi$ times the Noether charge associated with the Killing vector
whose norm vanishes on the horizon, providing that the Killing vector
is normalized to have unit surface gravity.  The normalization
condition effectively requires us to divide by the surface gravity
$\kappa$ of the black hole, so that \be S = {2\pi Q |_{horizon}\over
  \hbar \kappa} \, \, .  \ee A straightforward calculation \cite{domingo2}
gives the surface gravity of a black hole in generic dilaton gravity
to be: \be \kappa^2 \equiv - {1\over2} \nabla^\mu k^\nu \nabla_\mu k_\nu =
\left({V|_{horizon}\over 2}\right)^2  \,\, .\ee 
Moreover, on the horizon, according to
\re{Killing}), $| \nabla \Phi |^2 =0$, so that the final
expression for the entropy is: \be S = {4\pi \Phi |_{horizon} \0
  \hbar} \,\, . \ee
More generally, this procedure yields Eq.\
\re{thermo_entropy}).  For the specific case at hand
(i.e. $V(\Phi)
\propto \Phi = D(\Phi)$ and $Z=0$) $\Phi=B^3 =\sqrt{M}$ on-shell and
  \begin{equation} S_{\mbox{thermo}} \propto \sqrt{M} \, \, . \label{thermo}
 \end{equation}

We now  briefly comment on the Lorentzian signature case, in which the
coadjoint orbit will be non-compact and the total symplectic volume is
divergent.  Correspondingly, there will be infinitely many quantum
states leading to a meaningless $N=\infty$. Also there will be no
consistency condition enforcing a discrete spectrum for $M$; for any
value of $M$ there exists a Hilbert space (of the edge states) which
is infinite dimensional. We will  discuss this case further in what
follows. 

\section{The General Model}

We now apply the method of the
previous section to the generalized dilaton theories of Eq.\re{L}).  This
constitutes a qualitatively new step, as  these theories
do not in general
allow for a (standard) gauge theory formulation.
However, the notion of Poisson $\sigma$-models is applicable and
 will make these models tractable  in the present context.
For simplicity we will for the most part
restrict ourselves to the case $Z\equiv
0$, $D = \Phi$ in what follows.

The model \re{L}) may be described by \cite{PartI}: \be L[A_i,X^i] = \int_{\CM}
A_i \wedge dX^i + \2 \CP^{ij}(X(x)) A_i \wedge A_j \, \el PS with \be
\left(\CP^{ij}\right)(X) \equiv \left( \begin{array}{ccc} 0 &
  -V(X^3)/2 & -X^2 \\ V(X^3)/2 & 0 & \pm X^1\\X^2 & \mp X^1 & 0
\end{array} \right) \,\quad \, i,j \in \{ 1,2,3 \} \, , \el P where
the upper sign corresponds to the Euclidean, the lower sign to the
Lorentzian signature of the theory. In this context one can regard $A$ as
a triple of one-forms and $X$ as a triple of functions on spacetime.
This is in fact a generalization of the situation encountered in the BF-theory
\re{BF}) discussed above (with the obvious change of notation from $B$ to $X$).
The identification with geometrical variables is therefore as follows:
 The first
two components of $A$ coincide with the zweibein, the last one with
the spin connection. $X^1$ and  $X^2$ are Lagrange multipliers enforcing
zero torsion whereas $X^3 \equiv \Phi$. (For more general $D$ and/or $Z$ cf.\
\cite{PartI,PartIII}).

The Lie algebra of the gauge group is now replaced by the Poisson
bracket $\{X^i,X^j\} \equiv \CP^{ij}(X)$ on the target space $\dR^3$
spanned by the three linear target space coordinates
$X^i$.\footnote{$X^i(x)$ denotes the map from the spacetime $\CM$ to
  this 'target space', the space of values for the fields $X^i$.
  Although this space is $\dR^3$ and thus linear here, it carries a
  non-linear structure given by the two-tensor $\CP^{ij}$. For more
  details see \cite{ModLetts,Brief,PartI}, in particular the last
  two of them for a pedagogical introduction.} The latter reduces
to an ordinary Lie algebra, if $\CP^{ij}$ is linear in $X$. For
$\CP^{ij} = \varepsilon^{ij}{}_k X^k$ we recover the previous
$BF$-theory after a partial integration. In fact, the boundary term
picked up by the partial integration is nothing but the one that we
added in the course of our derivation of \re{deSnew}); thus \re{PS})
is already of the appropriate form for boundary conditions on $X$ ---
up to the gauge transformations, possibly, which will now be
discussed.

Making use of the central relation $(\partial \CP^{ij}/ \partial X^s)
\CP^{sk} + $ cycl.$(ijk) =0$ (the Jacobi identity for the bracket
$\{X^i,X^j\}$), it is easy to verify that under \be \d_\ep X^j =
\ep_i(x) \CP^{ij} \, , \quad \d_\ep A_i = d\ep_i + {\CP^{lm}}_{,i} A_l
\ep_m \,   \el symme the action \re{PS}) changes only by a total
divergence:
\be \d_\ep L = \int_{\6 \CM} X^i d \ep_i \, \, .  \el deltaepsilon
The symmetries \re{symme}) are an obvious generalization
of the standard nonabelian gauge symmetries (which arise for
linear
$\CP$).  By an appropriate choice of $\ep_i$
\re{symme})
may also generate diffeomorphisms: For $\ep_i
:= A_{i \mu} v^\m$ one verifies $\delta_\ep = \CL_v +$ (terms vanishing
by means of the field equations), which generalizes \re{diff}) to the
present context.

Following the first route for obtaining \re{CSnew}), we fix $X$ at
$\6 \CM$\footnote{This could be motivated by remarking that everything
written in the two paragraphs surrounding Eq.\ \re{casimir}) is valid here
too with the replacement $B \ra X$.} and add \re{deltaepsilon})
to \re{PS}), with $\ep_i$ now denoting additional fields, the 'edge
degrees of freedom'.
However, for $(X^i)|_{\6 \CM} := const$ the \re{deltaepsilon}) becomes a total
divergence and there is no action for the boundary degrees
of freedom. What did we do wrong? After all we know, e.g., that for
the special case \re{BF}) of \re{PS}) there does exist a nontrivial
action for the boundary variables.

The solution to this apparent puzzle stems from the fact that the symmetries
\re{symme}) are written in their {\em infinitesimal}\/ form, which
is not sufficient here. Upon inserting $g:=1+\epsilon(x)$ into
(\ref{coad}) and keeping only first order terms in $\epsilon$, one
also is left with a (meaningless) total divergence on $\6M$ only.
This illustrates the main difficulty in applying the approach of the
preceeding section to more general models; usually one knows the
symmetries generated by some constraints only in their infinitesimal
form, while the diffeomorphism group is difficult to handle because
it is
infinite dimensional and acts `nonlocally' on the fields. In the present
case  it is also far from trivial to `exponentiate' the local symmetries
\re{symme}).

At this point we apply a  trick that is standard in the
framework of Poisson $\sigma$-models: We change the parametrization of
the {\em target}\/ space of the theory and use coordinates on it which
are particularly adapted to the Poisson tensor $\CP$.
In particular, we  change  field variables $X^i \ra \widetilde X^i$
with
\be \widetilde X^i := (X^aX_a - \int^{X^3} V(z) dz\; ,\;
\left\{ \begin{array}{c} \arctan (X^2/X^1) \\ \mbox{artanh} (X^2/X^1)
\end{array} \right\} \; , \; X^3) \, \, \el CD
where $X^a X_a = \pm (X^1)^2 + (X^2)^2$ and
the upper/lower line of the second entry in \re{CD}) corresponds to the
upper/lower sign in \re{P}), i.e.\ to the signature of the theory. In
terms of the new variables,
the two-tensor $\CP$ takes the form
\be
\left(\CP^{\widetilde i \widetilde j}\right)(\widetilde X)
= \left( \begin{array}{ccc} 0 &
  0 & 0 \\ 0& 0 & 1 \\ 0&-1  & 0
\end{array} \right) \, \, , \el Ptilde
which simplifies \re{PS}) to
\be L =  \int_\CM \widetilde A_i \wedge d \widetilde X^i +
\widetilde A_2 \wedge
\widetilde A_3  \, . \el Ltilde
(Here we used the obvious target space covariance of \re{PS}).
In particular, $\widetilde A_i \equiv  (\partial X^j /\partial
\widetilde X^i )\, A_j$).

The parametrization $\widetilde X$, $\widetilde A$ of the field variables of
$L$ generically holds only on some local patch. This is, however,
sufficient for our purposes; there are further 'charts' on $\dR^3$
beside \re{CD}) bringing $\CP$ into the 'Casimir-Darboux-form'
\re{Ptilde}) and such charts may be patched together. In any of these
local charts the symmetries  (\ref{symme})  now take the form:
\be \delta \widetilde A_i = d \widetilde Y^i \;, \;\;\; \,
\delta \widetilde X^1 = 0 \,\; ,\; \;
 \delta \widetilde X^2 = \widetilde Y^3\,\; , \;\;
 \delta \widetilde X^3 = - \widetilde Y^2  \, \; . \el Y
Here the three fields $\widetilde Y^i$ may already be considered as
finite quantities, since in this form the symmetries are {\em
  linear}\/ in the parameter fields. Note that this form of the action and
symmetries even simplifies  the group theoretical  case \re{BF}).

Now we look at the (non-infinitesimal)
change of $L$ in this parametrization of the symmetries. Again it is a
surface term, but now one that is non-trivial:
\be \delta L = \int_{\6 \CM} \left(
  \widetilde{X}^i d \widetilde Y^i + \widetilde Y^2 d\widetilde Y^3
\right)\, .
\el delta
Given our boundary conditions on the fields $X$, we  may again drop the
first term in the above.  What we are left with
is the edge state action
\be L_{\mbox{symp}}\, [\widetilde Y_2,\widetilde Y_3]= \int_{\6 \CM}
\widetilde Y_2 \wedge d \widetilde Y^3  \,\, , \el symp
which is already of the simple form $\int p \dot q dt$ (after
identifying $\widetilde Y^3$ with a 'generalized coordinate' $q$ and
$\widetilde Y^2$ with a
'generalized momentum' $p$). Thus we did not have to find new
coordinates such as $\theta$ and $\varphi$ in \re{g}) in order to
obtain the point particle action at the boundary in Darboux-form; it
appeared in this form directly as a result of introducing coordinates
\re{CD}) adapted to the Poisson structure \re{P}).
It is also worth noting
 that it was the contribution of the $\widetilde A \wedge \widetilde A$
term that gave the nontrivial result \re{symp}) within the present
formulation; the surface term $-\int_{\6 \CM} \widetilde X^i \widetilde A_i$
did not contribute at all.

$L_{\mbox{symp}}$ is the analogue, or better generalization, of the
local $p \dot q$-form of the action \re{coad}). Once we are equipped
with some geometrical insight into the 'coordinates' $\widetilde X^i$,
we will be able to interpret $L_{\mbox{symp}}$ globally.  According to
\re{Ptilde}) it is $\widetilde X^1$, a Casimir coordinate of the
Poisson bracket on the target $\dR^3$, that determines the symplectic
leaves, as previously mentioned in the Introduction. Furthermore,
$\widetilde X^2$ and $\widetilde X^3$ serve as Darboux coordinates on
any of these leaves $\widetilde X^1 = const$. Note that, as an obvious
consequence of \re{Ltilde}), $\widetilde X^1(x)$ must also be a
constant on the spacetime. Thus the map from $\CM$ into the target
$\dR^3$ has to lie completely within a symplectic leaf. It turns out,
moreover, that the corresponding value of $\widetilde X^1$ on a
classical solution may be identified with the mass $M$, or at least a
function of $M$, whenever the notion of mass makes sense (cf., e.g., 
\cite{Steve}).

In view of the last two equations \re{Y}) it is now clear that the
fields $\widetilde Y^2$ and $\widetilde Y^3$ should take values not
just in an $\dR^2$, as they do locally, but in the same space as the
Darboux coordinates mentioned above. That is, they should also be
regarded as local coordinates of a symplectic leaf.  This symplectic
leaf is a {\em copy}\/ of the one of described by $\widetilde X^2$ and
$\widetilde X^3$, which is determined by the value of the Casimir
$\widetilde X^1 \sim M$. Similary one can introduce three linear
coordinate fields $Y$ and describe the symplectic leaf under
consideration by \be Y^a Y_a - \int^{Y^3} V(z) dz = M \el M where $M$
is the mass specified by the boundary conditions on $X$.

We  are now  in a position to  present the
correct  total gauge-invariant  action for the general nonlinear system
\re{PS}):
\be \widetilde L \, [A,X,Y] = L[A,X] +  L_{\mbox{symp}}[M[X],Y]  \,
\, ,\el total
where the fields $Y \in \dR^3$ are subject to \re{M}).
It is the appropriate generalization of \re{deSnew}). The  coadjoint
orbit encountered in \re{coad})
is generalized to a symplectic leaf. A (target space) coordinate
independent description of $L_{\mbox{symp}}$ reads \be L_{\mbox{symp}}
= \int_{\6 \CM} d^{-1} \Omega|_M  \, \, , \el indep
where $\Omega$ denotes the symplectic two-form on the symplectic leaf
specified by the value of $M=\widetilde X^1$ (in the above coordinates
$\Omega = d \widetilde Y^2 \wedge d \widetilde Y^3$). More precisely,
$\Omega$ is the symplectic two-form of  the Poisson bracket
\re{P}) with $X \ra Y$ and $Y \in \dR^3$ subject to \re{M}).

The symbol $d^{-1} \Omega$ has been used to denote a symplectic
potential (one-form), such as the one used in \re{symp}).
However, if the second homotopy group
 $\pi_2$ of the target leaf is not trivial, such a
potential will exist  only locally. But the action should be
independent of the particular local trivialization. The situation is
very similar to the WZNW-action \re{wznw});
the second term in \re{total})  may also  be
written as $(k/12\pi)\int_{\6 \CM} d^{-1} \left( dg g^{-1}
\right)^3$, and  single-valuedness of the quantum
action (path integral) restricts the ``coupling constant''. Here the role
of the latter is played by the mass $M$. For example, suppose $\6 \CM = S^1$ in
the
present case. Then the right hand side of \re{indep}) can be replaced by
$\int_{\CM} \Omega|_M$, which is independent (up to multiples of $2 \pi$)  of
how the integral  is continued from its boundary values if and only if
\be \oint \Omega|_M = 2 \pi n \hbar \, , \el Mquant $n \in Z$.
For $\pi_2$(leaves) $\neq 0$ this restricts the possible values of
$M$, a mechanism that we encountered already in a particular case when
dealing with the BF-theory.

Eq.\ \re{Mquant}) can also be obtained as a necessary and sufficient
condition for the geometric quantization of the phase space \re{M})
with symplectic form $\Omega |_{M}$ (cf.\ the Appendix). Precisely the
same condition arises in the Dirac quantization of the {\em bulk}\/
action as the {\em global}\/ integrability condition of the quantum
constraints, whose local integrability is ensured from the closure of
the quantum constraint algebra \cite{ModLetts}.  The question of
continuity or discreteness for the spectrum of $M$ is reduced to a
simple, purely topological question: The mass spectrum is discrete,
{\em iff}\/ the two-surfaces generated by \re{M}) have non-trivial
second homology.  Alternatively: A discrete spectrum of $M$ (within
the range of values $[a,b]$) occurs, {\em iff}\/ all the two-surfaces
(within the range $M \in [a,b]$) are compact.  In view of \re{M}) it
is easy to see that a discrete spectrum can occur only for the
Euclidean signature theory, but even in that case it is the 
exception.\footnote{We remark that in the approach of \cite{Barvinsky}
one obtains a purely discrete
mass spectrum for {all}\/ theories \re{L}) with Euclidean signature.}

At this point let us remark that for the case of a non-trivial first
homotopy of the surfaces \re{M}) of constant $M$, we believe that 
the above picture has to be corrected somewhat. 
The phase space of the edge particles
should then rather be regarded as the universal covering of the level
surfaces of $M$
(with the symplectic form given by the pull-back of $\O$ under the
covering map). The origin for this slight modification comes from
large gauge transformations and is basically the same as the one that,
in the special case of Lorentzian deSitter gravity, makes
$\widetilde{SO_e}(2,1)$ the correct gauge group rather than
$SO_e(2,1)$.

%This occurs,
%e.g., already in the Lorentzian deSitter model. There the level surfaces
%of \re{M}) are hyperboloids, which are multiply connected for $M>0$.
%But since the correct gauge group for the Lorentzian deSitter  gravity
%is the universal covering group $\widetilde{SO_e}(2,1)$ we should look
%for the coadjoint orbits of this group, and that is nothing but the
%universal covering of the coadjoint orbits of $so(2,1)$, which are the
%above hyperboloids. This generalizes to all of the models \re{L}) with
%non-simply connected symplectic leaves.

We are left with the task of determining the number of quantum states
on $\6 \CM$. Here we will content ourselves with a number of general
statements and then briefly focus on spherically symmetric and string
inspired dilaton gravity for
illustration. The number
of quantum states of the mechanical model described by
$L_{\mbox{symp}}$ is finite, {\em iff}\/ the two-surface \re{M}) (its
phase space) is compact. As found above in this case the mass is
discretized. For any one of these allowed values, $M = M_n$ such that
\re{Mquant}) is satisfied, the number of quantum states may be approximated
very well by $n$ (as follows from the same argument used
in the preceeding section or by the more precise consideration in the
footnote of the Appendix). Still only in exceptional cases the
entropy obtained via $S=S_{st.m.} \propto \ln (n)$ will fit the
thermodynamical $S= S_{thermo}$ given in \re{thermo_entropy}).

Actually, we can be even more concrete on this point. A compact leaf
is obtained, iff there are two successive zeros of the function $M +
\int^{Y^3} V(u) du$ (as a function of $Y^3$) between which this
function is positive. (Note that due to \re{M}) this expression
equals $Y^aY_a = (Y^1)^2 + (Y^2)^2$). Denote the corresponding two
values of $Y^3$ by $(Y^3)_{>}$ and $(Y^3)_{<}$, respectively, (with
$(Y^3)_{>} > (Y^3)_{<}$).  According to our knowledge of the
symplectic form the integral over $\Omega$ is computed easily (just
use Darboux coordinates, cf.\ Eq.\ \re{CD})): \be \oint \O = 2 \pi
[(Y^3)_{>}-(Y^3)_{<}] \,\, . \el oint Together with \re{Mquant}) and
$S_{st.m.} \propto \ln n$ this then yields \be S_{st.m.} \propto \ln
\left([(Y^3)_{>}-(Y^3)_{<}]/\hbar\right) \el comp for the case of
 compact phase spaces. In general this
disagrees with
\re{thermo_entropy}). Agreement would be found only for the (almost
pathological) case $(Y^3)_{<} = - \exp\left((Y^3)_{>}\right)
+(Y^3)_{>}$ (at least if one takes serious \re{thermo_entropy}) also
for the case of several horizons).  In the instance of the previous
section we had $(Y^3)_{>} = -(Y^3)_{<} = \sqrt{M}$ and \re{comp})
reduces to \re{SdeS}), as it should.

In all the cases where the surfaces \re{M}) are non-compact there will
be an infinite number of edge states. At least in cases where these
surfaces have trivial topology $\dR^2$, there seems to be no sensible
way to extract a finite number from that quantization.  For phase
space topology $\dR^2$ we know that there exist coordinates $q,p$ on
phase space such that $\Omega = dp \wedge dq$ {\em globally}\/
(Darboux theorem). If, furthermore, the respective symplectic leaf is
complete with respect to the flow of Hamiltonian vector fields, this in
turn implies that the quantization of the system is unique up to
unitary equivalence. This is most easily seen by noting that then the
transition to coordinates $q,p$ maps the system under consideration to
the one of an ordinary point particle on the line.  So here we cannot,
for example, do as Carlip did in Lorentzian 2+1 gravity, and count the
number of irreducible representations of the operator algebra to be
quantized.  This kind of `regularization' would yield a meaningless
{\em one}\/ in the present context.

As an illustration of these statements let us finally specialize to
ordinary dilaton gravity (the so-called ``string-inspired'' model) as
well as to spherically reduced gravity \cite{spher}. Both of these
models are governed by an action of the type \re{L}), albeit not of
the particular form with, e.g., $Z \equiv 0$ to start with.  The
appropriate expressions replacing \re{M}) in these cases are \be
M_{\mbox{dil}} = {Y^a Y_a \0 4 \lambda} + \lambda Y^3 \; \; , \; \; \;
M_{\mbox{SS}} = - Y^a Y_a + \sqrt{Y^3} \, \, , \el surf where $\lambda
>0$ and $Y^3$ is restricted to positive values and. Clearly \re{surf})
picks out non-compact surfaces for Lorentzian signature, $Y^a Y_a = -
(Y^1)^2 +(Y^2)^2$ (as is always the case with expression of the form
\re{M})). Unfortunately, it does so here too for Euclidean signature,
$Y^a Y_a = (Y^1)^2 +(Y^2)^2$.  Moreover, for any value of the mass $M$
these surfaces are simply connected and diffeomorphic to $\dR^2$. In
the Euclidean case they are also complete.\footnote{For ordinary
  dilaton gravity, e.g., this is most obvious by noting that $Y^1$ and
  $Y^2/2\lambda^2$ are possible Darboux coordinates on $M_{\mbox{dil}}
  = const>0$ and that their range is all of $\dR^2$ on these
  surfaces.} Thus, for reasons discussed above, the Hilbert space of
the edge system coincides uniquely with the one of square integrable
functions on the line $\dR$. In the Lorentzian case the leaves are not
complete with respect to the flow of Hamiltonian vector
fields\footnote{$\partial/\partial Y^3$, e.g., is the Hamiltonian
  vector field of $\ln Y^+$ and thus is incomplete with respect to its
  flow parameter since $Y^3 > 0$ (on-shell $X^3$, of which $Y^3$ is a
  copy, equals the exponential of the dilaton field $\Phi$ or the
  square of the Schwarzschild coordinate $r$, for the string inspired
  and the Schwarzschild case, respectively).} (just as the
spacetime is not complete with respect to geodesic vector fields). 
There are, however, still infinitely many quantum states associated
with these phase spaces since the total symplectic volume diverges (in
the Lorentzian case the quantization is similar to the one of a
particle restricted to an interval).

Thus we get infinitely many edge states in the Lorentzian {\em and}\/
the Euclidean theories here.  As with most 2-D models, therefore, we 
cannot use  the present approach to calculate a statistical
mechanical entropy for these two prominent cases.

\section{Discussion and Outlook}

We have attempted to calculate the statistical mechanical entropy for
black holes in $2D$ generic dilaton gravity in terms of edge states that
arise when the event horizon is treated as a boundary to spacetime.
We discussed various routes suggested in the literature for obtaining
the action governing these boundary modes.

%%%%%% None of those seemed
%%%%%really compelling to us.
 Since none of these
seemed compelling to us,  we suggested a modified reasoning where
%however,
%%This `however' is needlessly provocative: in my opinion
%at least our modification leads to a nice physical picture--gabor
the boundary does not coincide with the event horizon,
but instead becomes dynamical.
In this picture the ``boundary degrees of
freedom'' are nothing but the vibrational modes of this boundary
surface.

We then determined the Hilbert space of the boundary states.  In
essentially all cases we found the number of states to be too few, or
too many, to provide a microscopic source of black hole entropy.

In the cases for which the edge state phase space \re{M}) is compact,
we generically obtained the logarithm of the value expected from various
thermodynamic considerations, cf.\ Eqs.\ \re{thermo_entropy}) and
  \re{comp}). Here we note again that the logarithmic result
  \re{comp}) seems in qualitative agreement with the numerical results
  obtained by Srednicki \cite{Sred}, who determines the entanglement
  entropy for a massless scalar field in spacetimes of various
  dimensions.  For 1+1 dimensions, he gets an  entropy proportional
to $ \ln{R}$,
  where, in his lattice calculation, $R$ was a radius midway between
  the outermost point traced over and the innermost point not traced
  over. However, the logarithmic result does not seem
  satisfy the simplest version of the first law of thermodynamics 
 $TdS=dE$, which is obeyed by
the Bekenstein-Hawking entropy.

On the other hand, if the phase space \re{M}) (or the universal
covering thereof) is not compact, there is no quantization condition
on the thermodynamical entropy (or mass) and a finite degeneracy
cannot be calculated by these methods.

In principle this infinity is less devastating than the
previous result, where generically
there were far too {\em few}\/ quantum states.
In the case of an infinite dimensional Hilbert space of the edge states
one might  still hope  for some additional constraint that
would restrict the number of {\em physical}\/ states to a finite, but
``large'', number. %?
One possibility, suggested to us by Carlip \cite{carlip_private}, is
to look for a constraint
that generates diffeomorphisms along the Killing direction. Such a constraint
would be interpreted as a ``remnant of the Wheeler-DeWitt constraint''
that needs to be imposed on physical states. Unfortunately,
a preliminary investigation suggests that  there is no non-trival action of
the generator of diffeomorphisms along the Killing vector
 on the edge degrees of freedom. The reasoning is as follows:
 As is obvious from
\re{Killing}), lines of constant $\Phi \sim X^3$ coincide with the
Killing lines on the spacetime. So the generator $k$ of the
diffeormorphisms into the Killing directions may not change $X^3$.
Since, furthermore, on-shell $M=X^aX_a - \int^{X^3} V(z) dz = const$,
according to \re{casimir}) also $X^aX_a$, the only Lorentz invariant
combination of $X^a$, is left invariant by $k$. Thus, up to possibly
Lorentz transformations (in the frame bundle), $X^i$ will be left
invariant by $k$. As a consequence also the local edge phase space
variables $\widetilde Y^2$ and $\widetilde Y^3$ cannot be changed by
$k$, cf.\ Eq.\ \re{Y}). As a result there is no non-trival action of
$k$ on the edge degrees of freedom.

Actually at the Hamiltonian level it is easy to see that the
appropriate generator in the phase space of the bulk action is nothing
but the Casimir or mass functional $\widetilde X^1 \sim M$. Thus the
``dynamics'' in the Killing direction resides precisely in the
``abelian part'' of the action, which was found to give a total
divergence contribution to the boundary action only, cf.\ Eqs.\
\re{delta},\ref{symp}) and \re{coad2}). This coincides also nicely
with the fact that the Hamiltonian of our boundary action was found to
vanish identically. From these considerations it appears that there
is no need for a ``remnant of the Wheeler-DeWitt constraint''; the
symmetries in the Killing direction were already factored out  at
the classical level when part of the action was found to drop out as a
total divergence.

Irrespective of these general considerations, the only candidate for
such a generator that we can think of would be a function of only
$Y^3$ (since $X^3 = const$ are the Killing lines supposedly at least
$Y^3$ should be left unchanged). But such an ansatz, considering
$Y^3=const.$ as an additional constraint, does not seem to produce
reasonable results either: In the example of the Lorentzian deSitter
model, where the Poisson algebra \re{P}) is linear, we may apply an
algebraic quantization procedure analogous to the one used in the last
subsection of the Appendix.  We then have to count the eigenstates of
the (hyperbolic) generator $Y^3$ in the $sl(2,\dR)$-representations.
In the principal continuous series this number is just two (two-fold
degeneracy) while in the highest and lowest weight representations it
is one only, cf., e.g., \cite{Wybourne}. This attempt therefore also
seems to fail.\footnote{In any case, the generator $Y^3$ corresponds
  to (non-compact) frame bundle ``rotations'' (Lorentz
  transformations) and not to these diffeomorphisms, which act
  trivially on the edge degrees $Y^i$, as argued above.}

%But let us ignore these general considerations for a moment and
%consider the result obtained when taking some other  reasonable
%generator on the symplectic leaf. A candidate for such a generator may
%be considered a function of $Y^3$ only, because, after all, $X^3 =
%const$ are the Killing lines and thus, supposedly, at least $Y^3$
%should be left unchanged. So, let us consider $Y^3=B^3=const.$ as a
%constraint on the quantum states of the Lorentzian deSitter model. For
%this purpose we may apply an algebraic quantization procedure
%analogous to the one used in the last subsection of the Appendix, as
%in the deSitter model the Poisson algebra \re{P}) is linear. We
%therefore have to count the eigenstates of the (hyperbolic) generator
%$Y^3$ in the $sl(2,\dR)$-representations. However, in the principal
%continuous series this number is just two (two-fold degeneracy) while
%in the highest and lowest weight representations it is one only, cf.,
%e.g., \cite{Wybourne}. This attempt therefore also seems to
%fail.\footnote{In any case, the generator $Y^3$ corresponds rather to
%  (non-compact) frame bundle ``rotations'' (Lorentz
%  transformations) and not to these diffeomorphisms, which act
%  trivially on the edge degrees $Y^i$, as argued for above.}

%%%%%%%%%%%%%%%%%%%%end of major change

It seems to us very important to be able to understand  why the method
of Carlip and Balachandran et al.\ appears to fail in the present
context.

Possibly a further analysis of the ``abelian part'' of the action,
$\int \widetilde A_1 d \widetilde X^1$, that was found to yield only a
total divergence contribution to the boundary action \re{symp}), could
provide some insight. Dropping the rest of the action, this comes down
to the study of an abelian $BF$-theory. For the latter the gauge
theoretic approach applied in this paper did not yield any boundary
modes, irrespective of the chosen boundary conditions. The Hamiltonian
approach presented in the Introduction, on the other hand, suggests
the existence of boundary modes, although their existence will depend
decisively on the chosen boundary conditions\footnote{In the context
  of, e.g., the Schwarzschild black hole the quantities $O_\xi$
  stemming from the abelian part of the action are {\em either}\/ to
  be identified with the mass $M$ and its conjugate ``time''-variable
  $T$, found in the last two references of \cite{spher}, {\em or}\/,
  for our boundary conditions, will be just quantities fixed on the
  boundary. Note also that, in the standard way of thinking about the
  Schwarzschild black hole, $M$ and $T$ are rather regarded as the
  bulk modes of the black hole.} As discussed above, furthermore,
$\widetilde X^1 \sim M$ is the generator of the Killing time
transformations on the Hamiltonian level.  Last but not least, as
remarked to us by Steve Carlip \cite{carlip_private}, {\em if}\/ we
regard the partition function of the boundary action \re{coad2}) of
Euclidean deSitter gravity {\em without}\/ an imaginary unit in front
of the action, then we may not drop the ``total divergence'' $\dot
\lambda$, stemming again from the abelian part of the bulk; rather,
maps $g(t)$ with a non-trivial winding number for $lambda$, will yield
a contribution that goes as the {\em exponential}\/ of $\sqrt{M}$. For
these reasons we believe that {\em if}\/ there is a remedy for the
statistical mechanical entropy of the general class of 2d models
considered in the present paper, it should come from a different
treatment of the abelian part of the action.

%One of the
%reasons why we suggest the study of this simplified problem stems from
%the observation that following the gauge theoretic strategy,
%recapitulated in the body of the paper, there then is {\em no}\/
%boundary action at all. The Hamiltonian argumentation presented in the
%Introduction, on the other hand, should yield non-trivial edge
%observables. [This is connected to the fact that the gauge groups
%considered in the Lagrangian and the Hamiltonian formulation can
%differ in the asymptotic. An example for this scenario is provided by
%the ``time''-variable $T$ of the Schwarzschild black hole conjugate to
%$M$ found in the last two references of \cite{spher}. 
%Note that both
%$M$ and $T$ are contained in the abelian part of the Schwarzschild
%action in its Poisson $\s$-form.  We also remark that 
% usually $M$ and $T$ are
%considered as {\em bulk}\/ degrees of freedom, while in the context of
%Eq.\ \re{O}) \cite{Balchandran} they are rather ascribed to the
%boundary]. Moreover, 
%$M \sim widetilde X^1$ is precisely the generator
%of the Killing time transformations on the Hamiltonian level discussed
%above. And, last but not least, 
%as remarked to us by Steve Carlip \cite{carlip_private}, 

In view of the preceeding remarks on the conceptual difficulties in
deriving the boundary action, on the other hand, it seems natural also
to give up the idea that the boundary action is determined uniquely by
the bulk action and the boundary conditions.  Perhaps one should be
more flexible about what boundary action to consider, a point of view
that seems quite close in spirit to the considerations of some recent
works of Balachandran et al. In the case of \re{L}) it may be that
some different boundary action than the one studied in the present
paper, which may be postulated rather than derived, would yield a
better agreement of $\ln N$ with the semiclassical result
\re{thermo_entropy}).

Last but not least we remark that in the treatment of the boundary
action we did not take into account any interaction with the bulk
degrees of freedom. In the language of Carlip the bulk variables
entered as ``external currents'' only. In particular the mass $M$
entering in the boundary conditions was  non-dynamical;
one quantized  the edge degrees on a background spacetime of fixed mass
$M$. In this sense the present calculation was semiclassical only.
It may also well be that a combined treatment of the coupled
bulk-boundary system would yield more reasonable results.

We close with a final word on the relationship between Carlip's
calculation and the
string theoretic calculations of the statistical mechanical entropy of
black holes.  Clearly neither method is completely successful:  so far,
Carlip's edge state method only seems to work for the 3-d BTZ
(extremal and non-extremal) black holes;  the stringy calculations yield
the Bekenstein/Hawking entropy for certain 4 and higher dimensional extremal
and near-extremal black holes.  Furthermore, the stringy method appears
to have relation with the spacetime geometry.  Perhaps by understanding
the relationship between the two methods for extremal 3-d black holes, we
could understand the statistical mechanical origin of the higher dimensional,
and more physically realistic, non-extremal black holes.

\section*{Acknowledgement:}

The authors thank A.\ Barvinsky, S.\ Carlip, H.D.\ Conradi, D.\ 
Giulini, T.\ Jacobson, D.\ Marolf, and P.\ Schaller for useful
discussions. T.S.\ is grateful also to the Erwin-Schroedinger
Institute for hospitality during last summer.  This work was supported
in part by the Natural Sciences and Engineering Research Council of
Canada as well as by the Austrian FWF project 10.221-PHY.

\newpage
\begin{appendix}
\renewcommand{\theequation}{A.\arabic{equation}}
\setcounter{equation}{0}
\section*{Appendix: Counting States on the su(2)-coad\-joint orbit}

\par
\noindent
In the following we present three different methods for quantizing the
point particle action \re{Lcoad}) with $g \in SU(2)$. The first of
them is the most abstract one, but, in contrast to the successive two
methods,  it is this quantization procedure that
is applicable  in the more general context of quantizing
a general action of the form \re{indep}).

In particular, in all cases where the phase space of the point
particles governed by \re{indep}), i.e.\ the (connected components of
the) level surface \re{M}), is compact, the phase space is
diffeomorphic to a two-sphere. Moreover, there then always exist
spherical coordinates on the sphere such that the symplectic form $\O$
takes the form $\mbox{const.}(M) \, d (\cos \theta) \wedge d \varphi$.
Correspondingly, Eq.\ (\ref{Kaehler}) and everything derived from
this thereafter is valid in these cases too. In particular the compact
phase spaces are Kaehler and a holomorphic polarization yields
dim(Hilbertspace) $N= n+1$.  Only the relation between the integer $n$
and the mass $M$ will change from model to model, being determined by
the Bohr-Sommerfeld condition Eq.\ \re{Mquant}).

\medskip
\medskip
\noindent{\bf 1) Geometric Quantization}

\medskip
\par
\noindent
We will give a brief sketch of this
method; for more details see, e.g., \cite{Woodhouse}. In
the  main text [following Eq.\ (\ref{g})] we identified the symplectic
form as $\O =  \sqrt{M} d(\cos \theta) \wedge d\varphi$, where
$\theta$ and $\varphi$ are standard spherical coordinates on the
two-sphere. In geometric quantization $\hbar^{-1} \O$ becomes the
curvature form of a line bundle over the phase space, here $S^2$.
Such a bundle is
characterized by a winding number $n \in Z$, which may be thought of
as the homotopy class of the map from the 'equator' $S^1$ to
the structure group $C^* \sim U(1) \times \dR_+$, 
and which coincides precisely with
the Chern number $(1/2\pi) \, \oint \hbar^{-1} \O$ of the bundle.
This is the origin of \re{Mquant}), which lead to
(\ref{quant}) for the case of \re{Lcoad}).

The phase space $S^2$ is a Kaehler manifold. This is seen explicitly
by introducing a new complex coordinate $z=\cot (\theta/2)
\exp(i\varphi)$ ($Re(z)$ and $Im(z)$ are stereographic coordinates),
leading to \be \hbar^{-1} \O = i n \, \frac{dz \wedge d \bar z}{(1 + z
  \bar z)^2} \, , \el Kaehler where we replaced $\sqrt{M}$ by $n\hbar
/2$ already, which allows for the (local) Kaehler potential \be K=n \,
\ln \left(1 + z \bar z\right) \, , \el Pot where $\hbar^{-1} \O = i \6
\bar \6 K$.

The fastest way to determine the dimension of the Hilbert space
is to apply a version of the Riemann-Roch theorem, cf., e.g.,
\cite{Schlichenmaier}: The number of independent holomorphic sections
to a line bundle with Chern number $n \in \dN_0$ over a two-sphere
$S^2$ is $n+1$. This then equals the dimension of the Hilbert space
and thus coincides with the degeneracy $N$.

The above argument can be made more explicit:
In the patch of
applicability of the variable $z$ we choose a canonical potential $\alpha :=
-i \6 K$ ($\Rightarrow \hbar^{-1} \O = d \alpha$), and the
polarization vector $\6_{\bar z}$. The physical wave functions are then
seen to be holomorphic functions in this patch, for which we may
take as  a  basis
\be \Psi(z)= z^k \quad , \, \, k \in \dN_0  \, . \el Psiz
A complementary patch is provided by introducing the complex variable
$w = 1/z$ (stereographic coordinates with respect to the opposite pole). For
reasons of symmetry $\O$ has the same form in these coordinates,
\re{Kaehler}) with $w \lra z$. However, $K$ clearly changes in form
and so does $\alpha$:
\be \alpha = - in \6  \ln \left(1 + w \bar w\right)
+ d \left(\ln (w^{in})\right) \, \, . \el alpha
 Such an $\alpha$ has neither a $\bar z$
component nor a $\bar w$ component; since moreover
$\6_{\bar z} \propto \6_{\bar w}$, $\Psi$ is also holomorphic in the
$w$ coordinates. To have $\Psi$ fit together into a global holomorphic
section over all of the $S^2$, however, we learn from the second term in
\re{alpha}), which enforces an extra multiplicative 
gauge transformation  $\exp\left(-i \ln w^{in}\right) = w^{n}$ 
when switching the patches, that $\Psi(w) =
w^{n-k}$. This is non-divergent at $w=0 \leftrightarrow z = \infty$
only if $k$ is further restricted:
\be k=0,1,2, \, \ldots \, n  \, . \ee
This again leads to dim(Hilbert space) $=N=n+1$.

Two final remarks: First, the inner product between two states
$\Psi_1$ and $\Psi_2$ in the $z$-chart turns out to be $\langle \Psi_1
\mid \Psi_2 \rangle \propto \int \overline{\Psi_1(z)} \Psi_2(z) \left[
i dz \wedge d \bar z / (1+z \bar z)^{n+2} \right]$.  Second, in the
above we did not take into account a metaplectic correction, cf.\ 
\cite{Woodhouse}. This may lead to a slightly corrected
Bohr-Sommerfeld condition and, correspondingly, to a
slight shift in the spectrum for $M$ (cf.\ also the following two
subsections), which, however, is of little relevance for the present
considerations. 

\medskip
\medskip
\noindent{\bf 2) Elementary Oscillator Approach}

\medskip
\par
\noindent
We now wish to quantize \re{Lcoad}) by means of  an
elementary alternative method \cite{Proceedings}.
For this purpose we first parametrize $g$ by
\beq
g=\left(\ba{clcr} z_1\,\,&z_2\\-\bar z_2\,\,&\bar z_1\ea\right)\, ,
\eeq
where
\be
z_m = {1\over2}\left({q_m\over M^{1/4}}-i{p_m\over M^{1/4}}\right)
\quad , \, m = 1,2 \, ,
\ee
with the $p$'s and $q$'s real. In this parametrization, the action
 \re{Lcoad}) takes the simple form
\beq
S[q,p]=\oint_E dt\left(p_1\dot q_1+p_2\dot q_2\right) \, ,
\eeq
where we made use of the fact that $\mid z_1\mid^2+\mid z_2\mid^2=1$
due to  $g\in$ $SU(2)$. In terms of the phase space parameters
$p_m,\ q_m$
this last restriction on the variables $z_m$ becomes
a first class constraint, namely:
\be
{1\over 2}\left( (p_1)^2 + (p_2)^2 + (q_1)^2 + (q_2)^2\right)
\approx 2\sqrt{M}.\label{constraint}
\ee
This is of the form of the Hamiltonian of two coupled oscillators with
frequency 1 and total energy $2\sqrt{M}$. Moreover, the implementation of
this constraint in a Dirac procedure becomes just the stationary
Schroedinger equation for this auxiliary oscillator system. So here
 the standard result that the energy of oscillators becomes discrete on
the quantum level yields the quantization condition
\re{quant}) on $M$. Indeed $2\sqrt{M} = \hbar (n_1+n_2 +1)$, $n_m \in
\dN_0$ reproduces just this equation upon the substitution $n=n_1+n_2
+1$. In this approach the degeneracy becomes $n$.

Note that in the previous, geometric approach the spectrum for $M$ had
precisely the same form, except that $n \in {\dN}_0$ while here $n \in
\dN$. Shifting  $n \to n-1$ in the spectrum for $M$ of the previous
subsection  would yield complete agreement (including the degeneracy).

\medskip
\medskip
\noindent{\bf 3) Algebraic Approach}

\medskip
\par
\noindent
{}From the general considerations in the paragraphs
around Eqs.\ \re{M}) and \re{indep}) it is clear that the quantization
of the edge degrees of freedom of a general model described by a
Lagrangian \re{L}, \ref{PS}) comes down to the quantization of the
Poisson brackets
\be \{Y^i,Y^j\} = \CP^{ij}(Y) \el pbrackets
subject  to the constraint \re{M}):
\be Y^a Y_a - \int^{Y^3}  V(z) dz =  M \, \, , \el Mzwei
without which the bracket \re{pbrackets}) would be not
non-degenerate. In general the brackets \re{pbrackets}) will be
 highly non-linear and
an  algebraic mechanism of implementing them as operator
relations in some irreducible representation, respecting, furthermore,
the constraint \re{Mzwei}), will not be feasible. However, in
the present case of the interest in the Appendix,
(i.e. the edge
dymanics of (the $SU(2)$-version of) \re{BF})), the bracket
\re{pbrackets}) reduce to the linear $su(2)$ brackets
\be  \{Y^i,Y^j\} = \varepsilon(ijk) \, Y^k \el su2
with \re{Mzwei}) becoming $(\vec Y)^2 = M$. The algebraic quantization
of {\em these}\/ brackets is well-known certainly. The irreducible
representations are labelled by a half-integer valued spin $j$, they
have  dimension $2j +1$, and the spectrum of the Casimir $(\vec Y)^2$ is
$j(j+1)\hbar^2$.

The variables $\vec Y$ may be found also directly starting from
\re{Lcoad}). It is a nice exercise to verify that, using \re{g}),
\be -2 tr \left( \vec T g^{-1} T_3 g \right) =  \left(\ba{c}
\sin \theta \cos \varphi \\ \sin \theta \sin \varphi \\ \cos \theta
\ea\right) \, \, ,
 \el gY
where $\vec T$ may be represented by $-i \vec \sigma /2$, the
$\sigma_i$ being Pauli matrices.
Performing the change of variables from $g$ to $\vec Y$ defined by
$\vec Y := -2 \sqrt{M}  tr \left( \vec T g^{-1} T_3 g \right)$,
$(\vec Y)^2 = M$ is satisfied  by construction, and the brackets
\re{pbrackets}) follow immediately from the symplectic form $\O =
\sqrt{M} d\cos \theta \wedge d \varphi$,  obtained
from \re{Lcoad}) in the main text right after Eq.\ \re{g}).

Let us finally compare the results of the present subsection to those
of the preceding ones. For this purpose we first reformulate the
present results in terms of the integer valued variable $2j+1 =: n$:
\be M = {n^2 \hbar^2 \0 4} - {\hbar^2 \0 4} \quad, \, \, N = n \quad,
\, \, n \in \dN \, \, . \el final In this form it is obvious that the
difference between the algebraic approach and the previous oscillator
approach resides merely in a shift of the 'zero point energy' by $-
\hbar^2 / 4$. Agreement with the geometric approach is obtained by the
further shift $n \to n-1$ in the mass spectrum of subsection 1).
Moreover, we believe that a metaplectic correction to the geometric
quantization will produce at least the first of these two shifts,
possibly also the second one, which then would lead to complete
agreement with \re{final}).

\end{appendix}

\end{document}